\documentclass[aps,prb,twocolumn,showpacs,superscriptaddress,floatfix]{revtex4}
\usepackage{amsfonts}
\usepackage{graphicx}
\usepackage{amsmath}
\usepackage{times}
\usepackage{amssymb}
\usepackage[usenames]{color}
\usepackage[colorlinks,bookmarks=false,citecolor=blue,linkcolor=red,urlcolor=blue]{hyperref}

\newcommand{\su}[1]{SU($#1$)}

\begin{document}

\title{
Simplex solids in SU($N$) Heisenberg models on the kagome and checkerboard lattices
}

\author{Philippe Corboz}
\affiliation{Theoretische Physik, ETH Z\"urich, CH-8093 Z\"urich, Switzerland}
\author{Karlo Penc}
\affiliation{Institute for Solid State Physics and Optics, Wigner Research
Centre for Physics, Hungarian Academy of Sciences, H-1525 Budapest, P.O.B. 49, Hungary}
\affiliation{Department of Physics, Budapest University of
Technology and Economics and Condensed Matter Research Group of the Hungarian Academy of Sciences, 1111 Budapest, Hungary}
\author{Fr\'ed\'eric Mila}
\affiliation{Institut de th\'eorie des ph\'enom\`enes physiques, \'Ecole Polytechnique F\'ed\'erale de Lausanne, CH-1015 Lausanne, Switzerland}
\author{Andreas M. L\"auchli}
\affiliation{Institut f\"ur Theoretische Physik, Universit\"at Innsbruck, A-6020 Innsbruck, Austria}
\affiliation{Max-Planck-Institut f\"{u}r Physik komplexer Systeme, N\"{o}thnitzer Stra{\ss}e 38, D-01187 Dresden, Germany}

\date{\today}

\begin{abstract}
We present a numerical study of the SU($N$) Heisenberg model with the fundamental representation at each site for the kagome lattice (for $N=3$) and the checkerboard lattice (for $N=4$), which are the line graphs of the honeycomb and square lattices and thus belong to the class of bisimplex lattices. Using infinite projected entangled-pair states (iPEPS) and exact diagonalizations, we show that in both cases the ground state is a simplex
solid state with a two-fold degeneracy, in which the $N$ spins belonging to a simplex ({\it i.e.} a complete graph) form a singlet. Theses states can be seen as generalizations of valence bond
solid states known to be stabilized in certain SU(2) spin models.
\end{abstract}

%02.70.-c		:	Computational techniques; simulations
%71.10.Fd	:	Lattice fermion models (Hubbard model, etc.)
%03.67.-a		:	Quantum information
%67.85.-d,  ultracold Gases
%05.30.Fk, quantum statistical mechanics
%75.10.Jm Heisenberg model

\maketitle

%%%%%%%%%%%%%%%%%%%%%%%%%%%%%%%%
% WORD COUNT in V5
%%%%%%%%%%%%%%%%%%%%%%%%%%%%%%%%
%
%figure			aspect ratio	words
%states			871/378	->	85 words
%kagomemappings	749?/331	->	86 words
%res_su3kag_paper2	429/294 	->	123 words
%EDfigs_SU3		630/270	->	84 words
%res_su4cb_paper2	421/324	->	135 words
%P2Corrs_Checkerboard_SU4_N_20_border
%				555/226	->	81 words
%Total words in figures:			85+86+123+84+135+81=592		
%	
%Words in text					2940		
% (without abstract, title acknowledgments, references)		

%Total words 3007+592=3532
%maximally allowed: 3500 words

\section{Introduction}
The SU(2) Heisenberg model on the square lattice is one of the most studied systems in condensed matter physics, and its properties are by now well understood. Generalizations of this model to \su{N} with different values of $N$,
%different
lattice geometries, and
%also different
representations of \su{N} exhibit an extremely rich variety of different ground states. While such models have been the subject of many theoretical studies in the past decades,\cite{affleck1988-sun,marston1989,read1989,read1990,harada2003,kawashima2007,beach2009,hermele2009,hermele2011} they have recently attracted increasing interest thanks to the proposals to realize \su{N} symmetric Hubbard models  in experiments on ultracold fermionic atoms in optical lattices.\cite{wu2003,honerkamp2004,cazalilla2009,gorshkov2010}

%As in the \su{2} case, the \su{N} Heisenberg models can be derived from a generalized Hubbard model of $N$ flavors of fermions in a Mott insulating phase. Different numbers of particles per site correspond to different representations of \su{N}. In this work we focus on the case of exactly one particle per site, corresponding to the fundamental representation (a Young diagram with a single box) of \su{N} at each site.
%%In the terminology of Young tableaus this corresponds to one single box.
%%The Hamiltonian can then be written -->
% The low-energy effective model is given by the \su{N} Heisenberg Hamiltonian, that can be written
%in terms of a permutation operator $P_{ij}$ which exchanges the particles on neighboring sites,

As in the \su{2} case, the \su{N} Heisenberg models can be derived from a generalized Hubbard model of $N$ flavors of fermions in a Mott insulating phase. Different numbers of particles per site correspond to different representations of \su{N}. In this work we focus on the Heisenberg model with one particle per site, corresponding to the fundamental representation (a Young diagram with a single box) of \su{N} at each site.
%In the terminology of Young tableaus this corresponds to one single box.
%The Hamiltonian can then be written -->
The Hamiltonian can be written in terms of a permutation operator $P_{ij}$~\cite{toth2010} which exchanges the particles on neighboring sites,
\begin{equation}
\label{eq:H}
\mathcal{H}= \sum_{\langle i,j \rangle} P_{ij}. 
\end{equation}
The expectation value of the operator $P_{ij}$ is minimal (and equal to $-1$) when the wave function is antisymmetric on the $ij$ bond.
For \su{N}, a fully antisymmetric wave function can be constructed using exactly $N$ sites (and not only two sites as in the familiar \su{2} case). This is an \su{N} singlet, and the energy on any bond belonging to the singlet reaches
its lowest possible value $-1$.
%Consequently, the formation of an \su{N} singlet requires $N$ or a multiple of $N$ sites, and not only two sites as in the familiar \su{2} case.
This should be contrasted to the other class of \su{N} Heisenberg models,\cite{marston1989,read1989,read1990,harada2003,kawashima2007,beach2009} where conjugate representations on two different sublattices were used, and thus singlets can be formed between two sites for any $N$.

%%%%%%%%%%%%%%%%%%%%%%%%%%%%%%%%%%%%%%
%%%%%%%%%%%%%%%%%%%%%%%%%%%%%%%%%%%%%%%
\begin{figure}[]
\begin{center}
\includegraphics[width=1\linewidth]{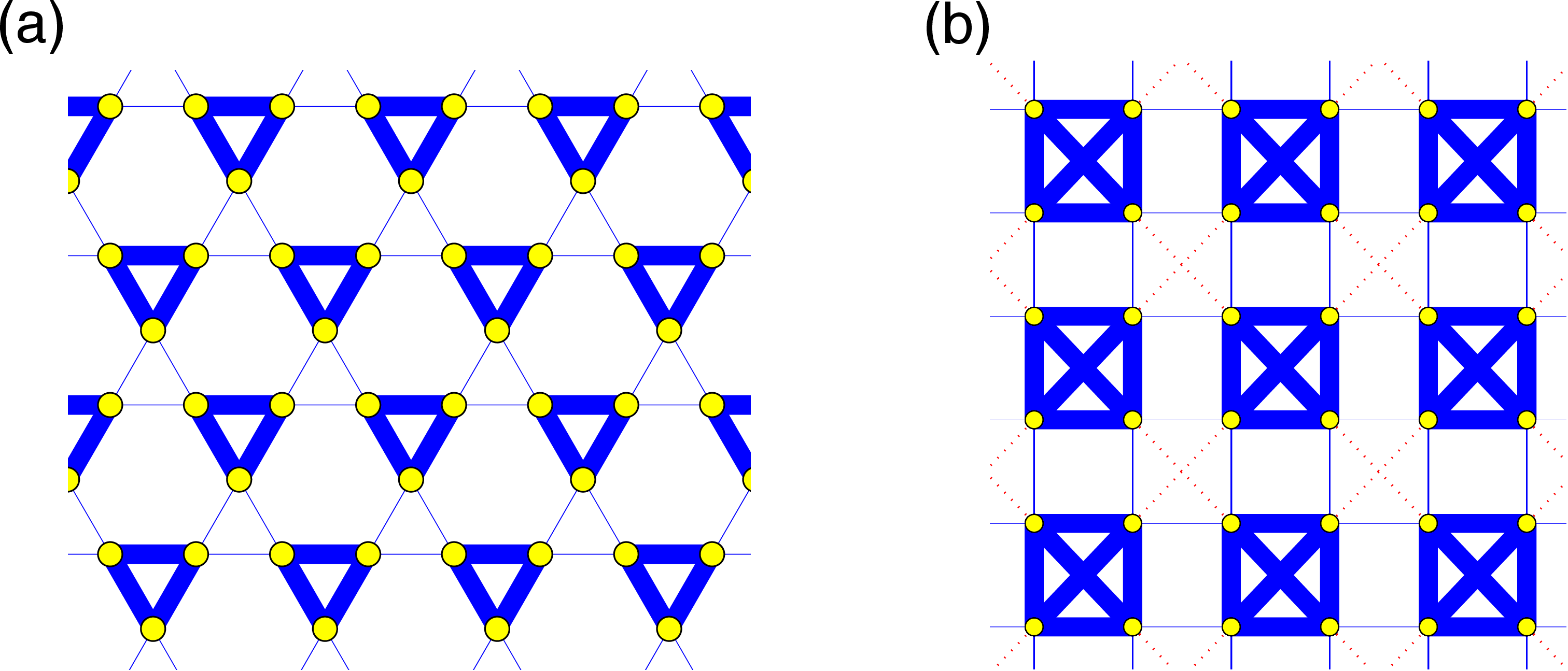}
\caption{(Color online) The simplex solid states obtained with iPEPS ($D=14$) for two different \su{N} Heisenberg models. The width of a bond is proportional to the magnitude of the bond energy, while blue (red dotted) bonds correspond to negative (positive) energies. (a) One of the two degenerate trimerized ground states of the \su{3} Heisenberg model on the kagome lattice. (b) One of the two quadrumerized ground states of the \su{4} Heisenberg model on the checkerboard lattice.
}
\label{fig:states}
\end{center}
\end{figure}
%%%%%%%%%%%%%%%%%%%%%%%%%%%%%%%%%%%%%%%
%%%%%%%%%%%%%%%%%%%%%%%%%%%%%%%%%%%%%%%

In this work we study the model on a particular class of lattices, so-called bisimplex lattices,~\cite{Henley2001} consisting of corner sharing simplices residing on an underlying bipartite lattice (equivalent to the line graph of the underlying bipartite lattice). In particular, we investigate the possibility of having an $N$-merized ground state,  i.e. where the lattice is covered by singlets which extend over $N$-site simplices.
\footnote{A $N$-site simplex is a cluster of $N$ sites in which each site is connected to all the other sites.}
%\cite{comment_simplices}
%
Such \emph{simplex solid states}\footnote{We have kept the terminology
simplex solid introduced by Arovas in Ref. \onlinecite{arovas2008}. However, since a lattice
symmetry is broken in the ground state, leading to a two-fold degeneracy, it is better seen as a generalization of a valence bond
crystal in the terminology of SU(2) valence bond states. In Ref.~\onlinecite{hermele2011} these states were called {\em valence cluster states}.}
%\cite{comment_simplex_solid} 
can be seen as a generalization of the valence bond solids (or valence bond crystals) known from certain \su{2} models,  as for example the dimerized state in the Majumdar-Gosh model.
Examples we consider here are the kagome lattice with 3-site simplices (triangles), and the checkerboard lattice with 4-site simplices (crossed-squares in Fig.~\ref{fig:states}(b), which are topologically equivalent to tetrahedra). Because of the underlying bipartite lattice (honeycomb for kagome, square for checkerboard) there are two possible ways to cover the lattice with singlets leading to a two-fold ground state degeneracy.

The occurrence of a symmetry broken state with a two-fold degeneracy on these lattices has already been observed in Refs.~\onlinecite{indergand2006,indergand_checkerboard} for a $t$-$J$ model away from half filling, which showed that this type of symmetry breaking is a likely candidate on these lattices. In Ref.~\onlinecite{arovas2008} Arovas derived parent Hamiltonians for the exact \su{N} simplex solid states on the kagome (for $N=3$) and the checkerboard lattice (for $N=4$). It is conceivable that these parent Hamiltonians are adiabatically connected to the \su{N} Heisenberg model, in the same way as the Affleck-Kennedy-Lieb-Tasaki (AKLT) state~\cite{affleck1987} is adiabatically connected with the ground state of the $S=1$ spin chain.
Simplex solid states on bisimplex lattices have also been predicted in Ref.~\onlinecite{hermele2011} based on studies of a particular large $N$ limit, with representations labelled by Young tableaus with $m$ rows and $n_c$ columns, where $N/m=k$ and $n_c$ are held fixed.

We use infinite projected entangled-pair states (iPEPS) and exact diagonalization (ED) to study the ground state of the $N=3$ case on the kagome and the $N=4$ case on the checkerboard lattice. 
%iPEPS is a (variational) tensor network ansatz for wave functions in two dimensions in the thermodynamic limit, which has  previously been used to determine the ground state of the square-lattice \su{4} Heisenberg model\cite{corboz11-su4} and the triangular- and square-lattice \su{3} Heisenberg model.\cite{bauer2011-su3} 
Both methods consistently predict a two-fold degenerate simplex solid ground state for both models, summarized in Fig.~\ref{fig:states}.

%outline
The paper is organized as follows: In Sec.~\ref{sec:iPEPS} we provide details on the iPEPS simulations, in particular, how the models are simulated using a square-lattice iPEPS. In Secs.~\ref{sec:kagome} and \ref{sec:checkerboard} we present the iPEPS and ED results obtained for the kagome and the checkerboard model, respectively. Finally, Sec.~\ref{sec:conclusion} summarizes our findings.

%%%%%%%%%%%%%%%%%%%%%%%%%%%%%%%%%%%%%%%
%%%%%%%%%%%%%%%%%%%%%%%%%%%%%%%%%%%%%%%
\begin{figure}[]
\begin{center}
\includegraphics[width=8.8cm]{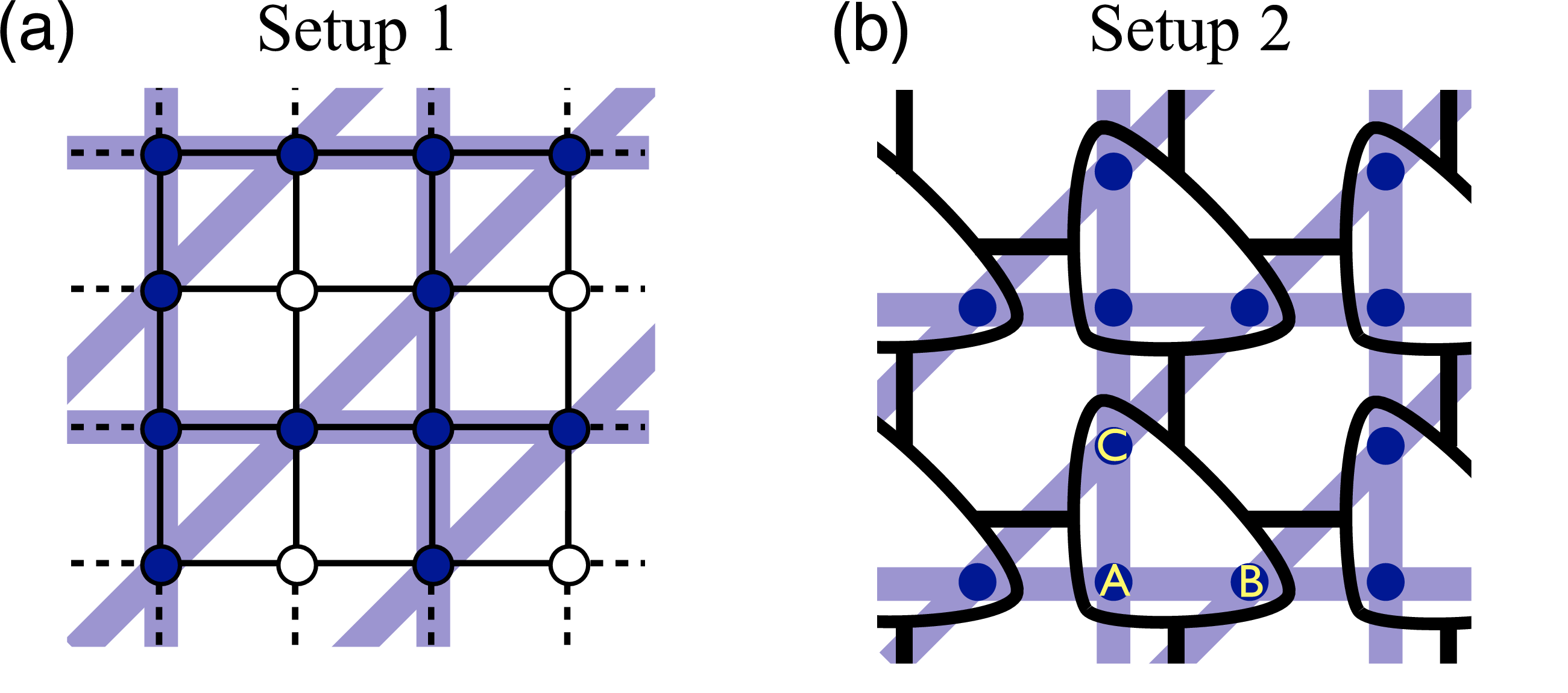}
\caption{(Color online) The two different simulation setups used to simulate the kagome lattice with the iPEPS method developed for the square lattice. Black circles and triangles correspond to tensors, black lines to connection between tensors (the physical index of a tensor is not shown). Interactions between physical sites (filled circles) are given by thick shaded lines. (a) Auxiliary tensors (white circles) are introduced to create a square lattice iPEPS.  The interactions along the horizontal and vertical direction correspond to nearest-neighbor couplings on the square lattice, whereas the interactions along the diagonal are treated as next-nearest neighbor interactions. (b) Three physical sites $A$, $B$, $C$ on the kagome lattice, each having a local dimension $d=3$,  are mapped into a block site with a local dimension $\tilde d=27$.
}
\label{fig:kagomemappings}
\end{center}
\end{figure}
%%%%%%%%%%%%%%%%%%%%%%%%%%%%%%%%%%%%%%%
%%%%%%%%%%%%%%%%%%%%%%%%%%%%%%%%%%%%%%%

%%%%%%%%%%%%%%%%%%%%%%%
% Description of the method used
%%%%%%%%%%%%%%%%%%%%%%%

\section{Infinite projected entangled-pair states (iPEPS)}
\label{sec:iPEPS}
%An iPEPS is a tensor network ansatz enabling an efficient representation of two-dimensional ground state wave functions in the thermodynamic limit.\cite{sierra1998, nishino1998, verstraete2004, nishio2004, murg2007, jordan2008} It consists of a unit cell of rank-5 tensors, with one tensor per lattice site, which is periodically repeated on the lattice. The first index of a tensor carries the local Hilbert space of the corresponding lattice site, the four remaining indices - the auxiliary bonds - connect the tensor to its four nearest neighbors (on the square lattice).
%The accuracy of the ansatz can be  controlled by the so-called bond-dimension $D$ of the auxiliary bonds. A bond-dimension $D=1$ simply corresponds to a product state (a site-factorized wave function), and upon increasing $D$ quantum fluctuations (entanglement) can be taken into account in a systematic way.
%-->
A projected-entangled pair state (PEPS) is an efficient variational ansatz for two-dimensional ground state wave functions.~\cite{sierra1998, nishino1998, verstraete2004, nishio2004, murg2007} It can be seen as a natural extension of a matrix product state (MPS), the underlying ansatz of the famous density matrix renormalization group (DMRG) method.~\cite{white1992}  The main idea is to represent a wave function by a trace of a product of tensors, with one tensor per lattice site. On the square lattice each tensor $T^p_{ijkl}$ has five indices: one index $p$ which carries the local Hilbert space of a lattice site with dimension $d$, and four indices $i,j,k,l$ - the auxiliary bonds with bond dimension $D$ - which connect to the four nearest-neighbor tensors. Thus, each tensor contains $dD^4$ variational parameters, and by varying $D$ the accuracy of the ansatz can be systematically controlled. A bond dimension $D=1$ simply corresponds to a product state (a site-factorized wave function), and upon increasing $D$ quantum fluctuations (entanglement) can be taken into account in a systematic way. 

Details on the iPEPS method for the square lattice can be found in Refs.~\onlinecite{jordan2008, corboz2010, corboz2011}, in particular how to \textit{optimize} the tensors (i.e. finding the best variational parameters) and how to compute expectation values by \textit{contracting} the tensor network (i.e. computing the trace of the product of all tensors). We performed similar iPEPS simulations already for the \su{4} Heisenberg model\cite{corboz11-su4} and the triangular- and square-lattice \su{3} Heisenberg model.\cite{bauer2011-su3} 

For the experts, we note that the optimization is done through an imaginary time evolution using the simple update,\cite{vidal2003-1, orus2008, jiang2008, corboz2010} and we have verified some simulation results also with the full update.\cite{corboz2010} The corner-transfer matrix method~\cite{nishino1996, orus2009-1,corboz2010} is used to  contract the tensor network, where the  error of the approximate contraction can be controlled by the boundary dimension $\chi$. The simulation results in this work are extrapolated in $\chi$, where the extrapolation error is small compared to the symbol sizes. To improve the efficiency of the simulations we used tensors with $\mathbb{Z}_q$ symmetry.\cite{singh2010,bauer2011}

%SIMULATION SETUPS
To simulate the checkerboard model we use the usual square lattice iPEPS ansatz where we treat the diagonal couplings as next-nearest neighbor interactions as described in  Ref.~\onlinecite{corboz2010-nn}.

For the kagome lattice we implemented two different simulation setups based on a square lattice iPEPS, which have the advantage that existing algorithms for the optimization and contraction can be reused.
For the first variant we use one tensor per lattice site, plus additional auxiliary tensors which are inserted to form a square lattice iPEPS, as sketched in Fig.~\ref{fig:kagomemappings}(a). [The bond dimension of the auxiliary tensors can be chosen as $D=1$ since all correlations are carried by the tensors on the physical sites.]
The couplings along the horizontal and vertical direction correspond to nearest-neighbor couplings between two tensors, whereas the remaining bonds necessary to form the kagome lattice are represented by the next-nearest-neighbor bonds in the square lattice, which can be treated as explained in Ref.~\onlinecite{corboz2010-nn}.
%
%whereas the Hamiltonian terms along the diagonal are implemented as next-nearest neighbor interactions. %
%
%
In the second setup we map the kagome lattice onto a square lattice by blocking three sites as illustrated in Fig.~\ref{fig:kagomemappings}(b). The original Hamiltonian is mapped onto a square lattice Hamiltonian with nearest-neighbor interactions between the block sites (see supplementary material). We point out here that we do not block three sites belonging to a triangle in the kagome lattice, since this would automatically bias the solution towards a trimerized state.

%OBSERVABLES
Since we work directly in the thermodynamic limit, the ground state wave function may exhibit spontaneously broken symmetries. We measure the energy on each bond, $E_b$, in the unit cell. If the energies are not equal on all symmetry related bonds, i.e. if the difference
\begin{equation}
\Delta E = \max(E_b) - \min(E_b)
\end{equation}
is finite, then the state breaks some lattice symmetries.

Furthermore, we compute the local ordered moment $m$ on each site,
\begin{equation}
\label{eq:m}
 m=\sqrt{\frac{N}{N-1} \sum_{\alpha,\beta} \left(\langle S_\alpha^\beta \rangle - \frac{ \delta_{\alpha\beta}}{N } \right)^2},
\end{equation}
where $S_\alpha^\beta = |\alpha\rangle\langle \beta|$ are the generators of \su{N} and $\alpha,\beta$ run over all flavor indices. A finite $m$ implies that the \su{N} symmetry is broken.

%We tried different unit cell sizes in iPEPS to test for possible stable color-ordered solutions, however all simulations consistently lead to an $N$-merized ground state. The  results presented in the following are obtained with a $2\times2$ unit cell of tensors.
%-->
We tried different unit cell sizes up to size $6\times 6$ in iPEPS to test for possible stable color-ordered solutions,  however all simulations consistently lead to an $N$-merized ground state. The  results presented in the following are obtained with a $2\times2$ unit cell of tensors.

%in particular a $6 \times 6$ unit cell which can accommodate the $\sqrt{3}\times \sqrt{3}$ color ordered state predicted by linear flavor-wave theory (see Sec.~\ref{sec:conclusion}) on the kagome lattice. We found that this state only appears as a metastable state in iPEPS, i.e., the $N$-merized ground state has a lower variational energy. The results presented in the following are obtained with a $2\times2$ unit cell of tensors.

%%%%%%%%%%%%%%%%%%%%%%%%%%%%%%%%%%%%%%%
%%%%%%%%%%%%%%%%%%%%%%%%%%%%%%%%%%%%%%%
\begin{figure}[]
\begin{center}
\includegraphics[width=1\linewidth]{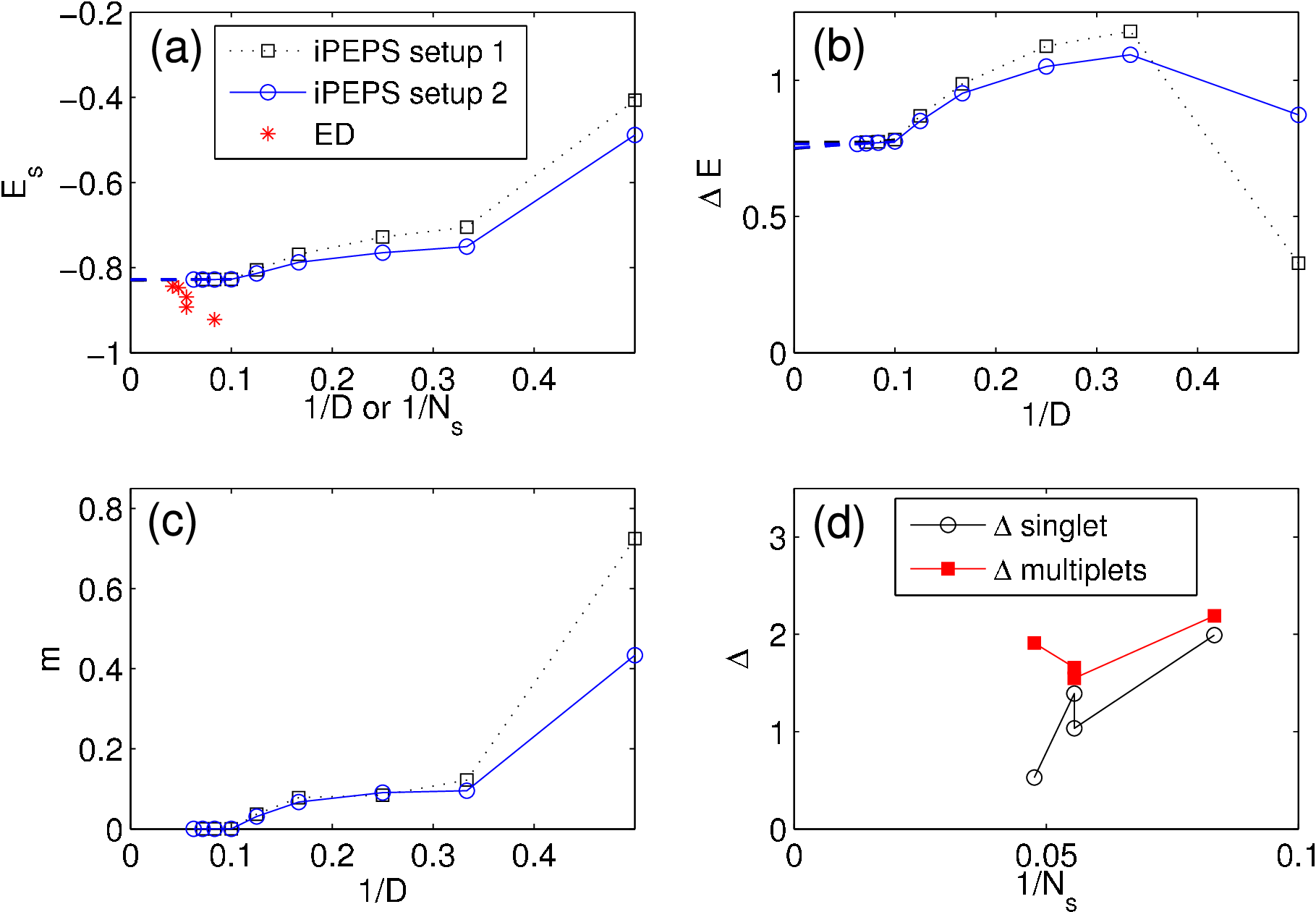}
\caption{(Color online) (a-c) iPEPS simulation results for the \su{3} Heisenberg model on the kagome lattice as a function of inverse bond dimension $1/D$ for the two simulation setups (cf. Fig.~\ref{fig:kagomemappings}). Extrapolations (dashed lines) are a guide to the eye. (a) Energy per site compared with the results from ED (plotted as a function of $1/N_s$). The estimated value in the infinite $D$ limit is $-0.829(1)$.
(b)~The difference in bond energies $\Delta E$ remains finite in the infinite $D$ limit, which shows that the state breaks translational invariance as illustrated in Fig.~\ref{fig:states}(a).
(c) Local moment which is strongly suppressed with increasing bond dimension. 
(d) ED results for the energy gaps to the first singlet and the first \su{3} multiplet excitations
plotted as a function of the inverse number of sites $1/N_s$.
}
\label{fig:su3kag}
\end{center}
\end{figure}
%%%%%%%%%%%%%%%%%%%%%%%%%%%%%%%%%%%%

%%%%%%%%%%%%%%
%RESULS SU(3) KAGOME
%%%%%%%%%%%%%%
\section{The \su{3} model on the kagome lattice}
\label{sec:kagome}
We consider the model \eqref{eq:H} on the kagome lattice for $N=3$ with the  fundamental representation at each lattice site. Singlet formation most naturally occurs  between three sites on a triangle, resulting into two possible coverings on the kagome lattice in which either all triangles pointing upwards, or all triangles pointing downwards form singlets. The latter case is illustrated in Fig.~\ref{fig:states}(a).

Figure~\ref{fig:su3kag}(a) shows the iPEPS energy per site, $E_s$ obtained with the two different simulation setups (cf. Fig.~\ref{fig:kagomemappings}) as a function of bond dimension $D$. It is in general not known how quantities depend on $D$. Here we find that for $D>10$ the energy does not change much anymore upon further increasing $D$. As an estimate in the infinite $D$ limit we take the middle between the value at the maximal $D$ and the value obtained from a linear extrapolation in $1/D$ for the four largest $D$. %This estimate $E_s=-0.836(8)$ compares well with the finite-size ED data (red stars in \ref{fig:su3kag}(a)), where the energy tends to increase with system size.

From Fig.~\ref{fig:su3kag}(b), it is clear that the difference between the highest and lowest bond energy $\Delta E$ is finite for all values of $D$, which indicates that the ground state breaks translational symmetry.   The strength of the individual bond energies in the lattice can be seen in Fig.~\ref{fig:states}(a), where the thickness of a bond is proportional to the magnitude of the bond energy. The resulting pattern is clearly compatible with a trimerized state. The strength of the trimerization decreases with increasing $D$ (except for the lowest $D=2$), but it seems to remain finite around $\Delta E = 0.76(1)$ in the infinite $D$ limit.

Since the trimerized state does not break the \su{3} symmetry, the local ordered moment $m$ defined in Eq.~\eqref{eq:m} should vanish.
Figure~\ref{fig:su3kag}(c) shows that $m$ is strongly suppressed with increasing $D$. In both simulation setups $m$ vanishes completely for $D\ge10$.

We note that we have also tested a $6 \times 6$ unit cell which can accommodate the $\sqrt{3}\times \sqrt{3}$ color ordered state predicted by linear flavor-wave theory (see Sec.~\ref{sec:conclusion}). We found that this state only appears as a metastable state in iPEPS, i.e., the $N$-merized ground state has a lower variational energy. 

\begin{figure}
\begin{center}
\includegraphics[width=4cm]{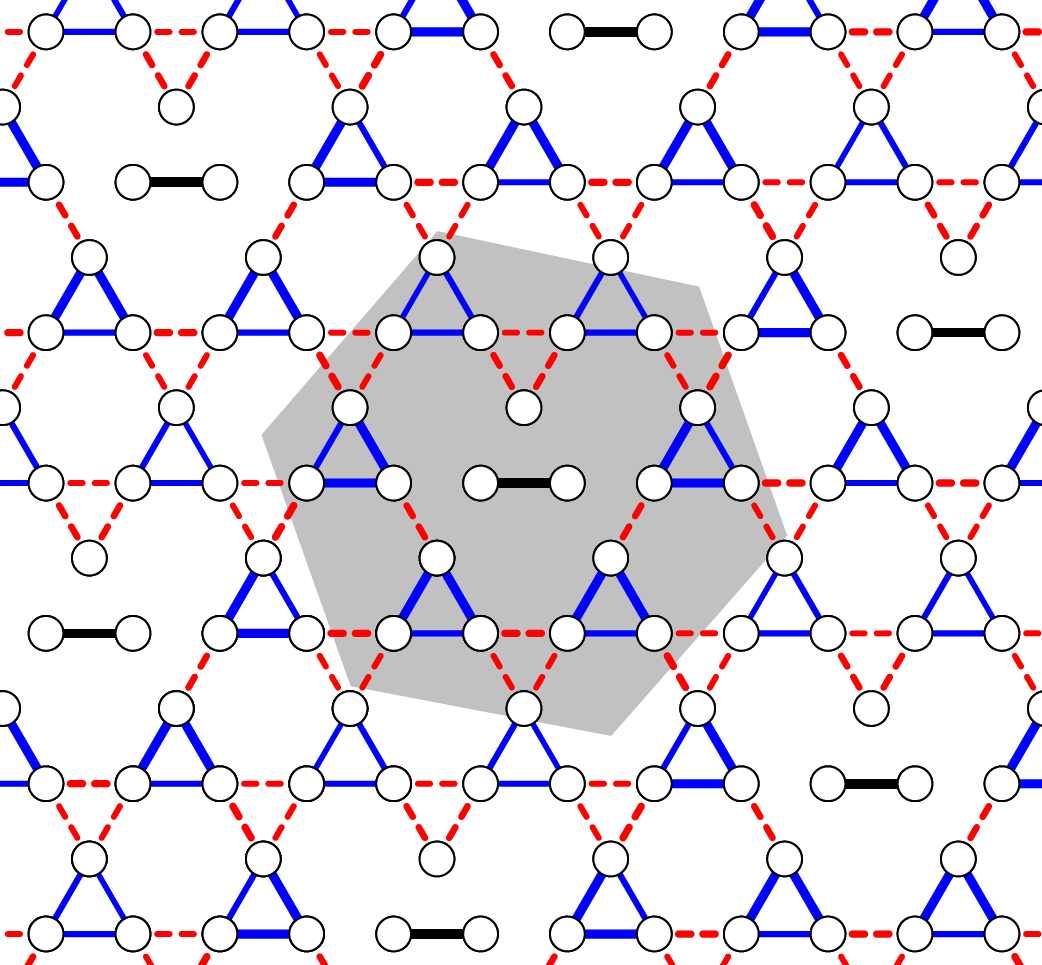} \hspace{0.4cm}
\includegraphics[width=4cm]{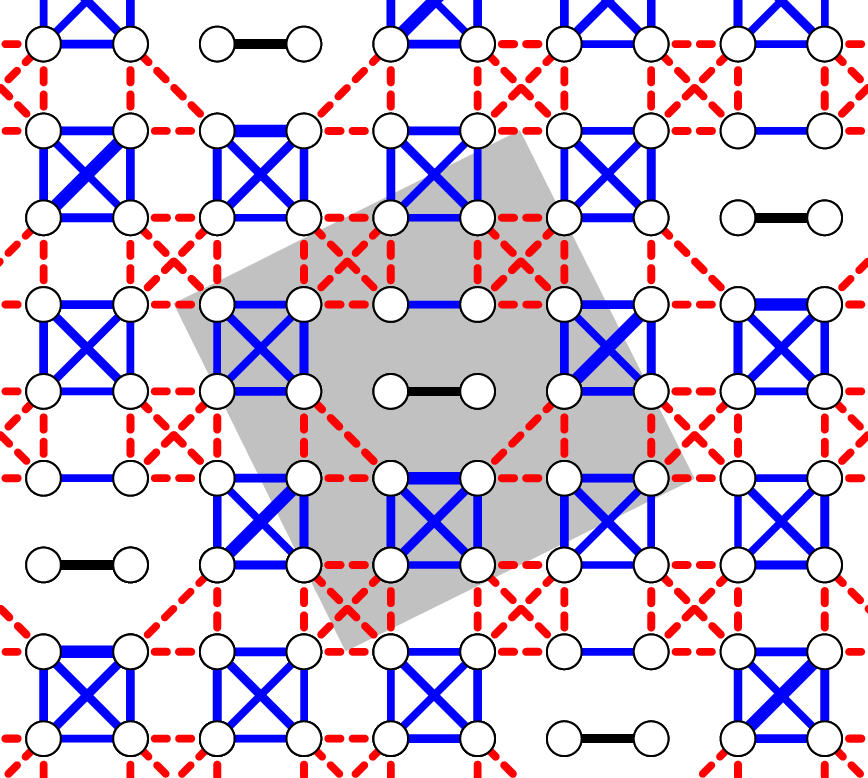}
\caption{(Color online) ED results for the connected bond energy correlations $\langle P_{ij}P_{kl}\rangle-\langle
P_{ij}\rangle\ \langle P_{kl}\rangle$. The reference bond is black, positive (negative) correlations are shown with blue (red dashed) lines. The
width of the lines is proportional to the correlation function. The periodic (Wigner-Seitz) cell is shaded in grey. Left panel: Results for the SU(3) Heisenberg model for a $N_s=21$ kagome sample. Right panel: Results for the \su{4} Heisenberg model on the checkerboard lattice for a $N_s=20$ sample.
}
\label{fig:EDresultsKagome}
\end{center}
\end{figure}

To further corroborate the iPEPS results we have performed ED simulations of the \su{3} Heisenberg model on finite size kagome samples of up to $N_s=24$ sites (c.f. Ref.~\onlinecite{aml_kagome} regarding the geometry of the clusters). The energies per site have been included in Fig.~\ref{fig:su3kag}(a) and the agreement between the two methods at large $D$ or $N_s$  is very good. In 
Fig.~\ref{fig:su3kag}(d)  we display two relevant energy gaps, the first one to the lowest singlet excitation at the $\Gamma$ point, and the second one to the first magnetic (i.e.~non-singlet)
excitation. While the singlet gap seems to collapse for larger $N_s$, in agreement with the lattice symmetry breaking scenario requiring a two-fold ground state degeneracy, the magnetic gap appears to stay finite, in agreement with the picture of a nonmagnetic singlet ground state. In the left panel of Fig.~\ref{fig:EDresultsKagome} we finally show the connected bond energy correlations (cf. caption), which
convincingly demonstrate the trimerization of the kagome lattice \su{3} Heisenberg model.

In summary, both iPEPS and ED provide strong indications that the ground state of the \su{3} Heisenberg model on the kagome lattice is trimerized.

%%%%%%%%%%%%%%%%%%%%%%%%%%%%%%%%%%%%%%%
%%%%%%%%%%%%%%%%%%%%%%%%%%%%%%%%%%%%%%%
%%%%%%%%%%%%%%%%%%%%%%%%%%%%%%%%%%%%%%
%%%%%%%%%%%%%%%%%%%%%%%%%%%%%%%%%%%%%%%
%%%%%%%%%%%%%%%%%%%%%%%%%%%%%%%%%%%%%%%
%%%%%%%%%%%%%%%%%%%%%%%%%%%%%%%%%%%%%%%

%\begin{figure}
%\begin{center}
%\includegraphics[width=1\linewidth]{P2Corrs_Checkerboard_SU4_N_20_border}
%\caption{(Color online) \su{4} Heisenberg model on the checkerboard lattice:
%ED results for the connected bond energy correlations $\langle P_{ij}P_{kl}\rangle-\langle
%P_{ij}\rangle\ \langle P_{kl}\rangle$ for a $N_s=20$  sample. The
%reference bond is black, positive (negative) correlations are shown in blue (red). The
%width of the lines is proportional to the correlation function.
%}
%\label{fig:EDresultsCheckerboard}
%\end{center}
%\end{figure}
%%%%%%%%%%%%%%%%%%%%%%%%%%%%%%%%%%%%%%%%
%%%%%%%%%%%%%%%%%%%%%%%%%%%%%%%%%%%%%%%

%DeltaE:0.71986 pm -0.046295

\section{The \su{4} model on the checkerboard lattice}
\label{sec:checkerboard}
The \su{4} model on the checkerboard lattice is described by the Hamiltonian \eqref{eq:H} with the fundamental representation of \su{4} at each lattice site. Thus the local dimension of a lattice site is four. %As mention in Sec.~\ref{sec:iPEPS} we treat the diagonal couplings as next-nearest neighbor interactions~\cite{corboz2010-nn}.

\begin{figure}[]
\begin{center}
\includegraphics[width=1\linewidth]{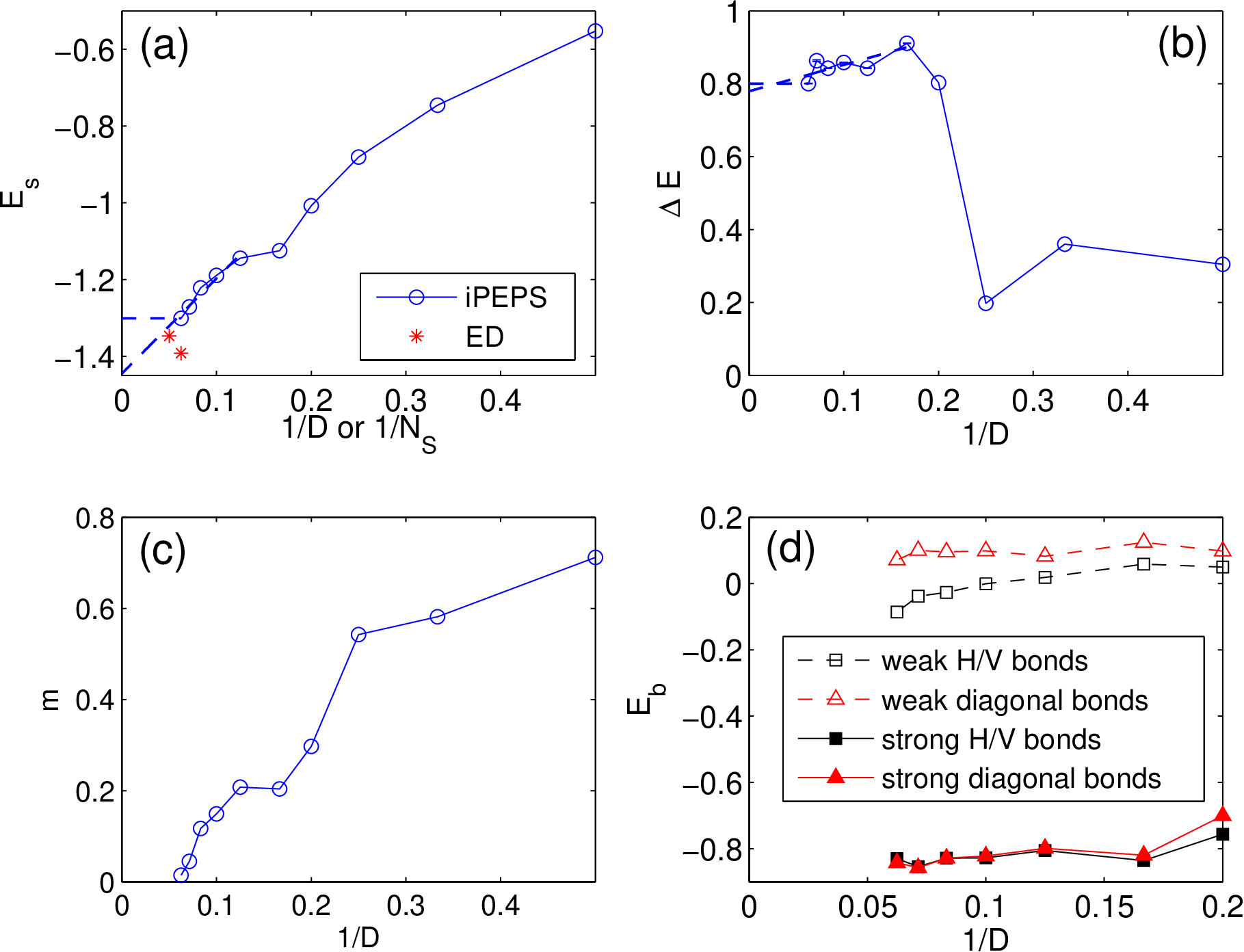} %does not scale to full linewidth?
\caption{(Color online)  iPEPS simulation results for the \su{4} Heisenberg model on the checkerboard lattice as a function of inverse bond dimension. (a) Energy per site, $E_s$, compared with the results from ED (plotted as a function of $1/N_s$).
(b) Different bond energies in the quadrumerized state. The strong horizontal and vertical bonds (H/V) have the same energy as the strong diagonal bonds for large~$D$.
(c) The difference in bond energies $\Delta E$ is finite which shows that the state breaks lattice symmetries [cf. Fig.~\ref{fig:states}(b)].
(d) The local moment decreases rapidly with $D$ and vanishes for large~$D$.
}
\label{fig:su4cb}
\end{center}
\end{figure}

Figure~\ref{fig:su4cb}(a) shows the iPEPS energies as a function of inverse bond dimension. We find that even for the largest values of $D$ used the energy still decreases considerably upon further increasing $D$. In the infinite $D$ limit we expect the energy to lie in the range $-1.45 <E_s<-1.30$. The ED result for finite systems $E^{N_s=16}_s=-1.392$, and $E^{N_s=20}_s=-1.347$ lie within this range.

The difference in bond energies $\Delta E$, shown in Fig.~\ref{fig:su4cb}(b), becomes large for $D\ge5$. Not all the weak bonds have exactly the same energy, see Fig.~\ref{fig:su4cb}(d). So we took  averaged energies of the weak and strong bonds to compute $\Delta E$. The distribution of the individual bond energies on the lattice can be seen in Fig.~\ref{fig:states}(b), clearly showing that quadrumers are formed. With increasing bond dimension $\Delta E$ slightly decreases but the data still suggest a rather large value in the range $\Delta E \approx 0.77-0.80$ in the infinite $D$ limit.

Figure~\ref{fig:su4cb}(c) shows that the ordered moment vanishes in the large $D$ limit, compatible with a quadrumerized state which does not break \su{4} symmetry.

Finally, ED results for the connected bond energy correlations are shown in the right panel of Fig.~\ref{fig:EDresultsKagome} for a $N_s=20$ sample. They further highlight the strong quadrumerization instability of the checkerboard lattice \su{4} Heisenberg model.

%%%%%%%%%%%%%%%%%%%%%%%%%%%%%%%%%%%%%%%
%%%%%%%%%%%%%%%%%%%%%%%%%%%%%%%%%%%%%%%
%%%%%%%%%%%%%%%%%%%%%%%%%%%%%%%%%%%%%%%
%%%%%%%%%%%%%%%%%%%%%%%%%%%%%%%%%%%%%%%
%%%%%%%%%%%%%%%%%%%%%%%%%%%%%%%%%%%%%%
%%%%%%%%%%%%%%%%%%%%%%%%%%%%%%%%%%%%%%%%

\section{Discussion}
\label{sec:conclusion}
The iPEPS and exact diagonalization results reported in this paper give clear evidence in favor of the stabilization 
of simplex solids for \su{3} on kagome and for \su{4} on the checkerboard lattice. The underlying mechanism is 
the formation of SU(N) singlets on the simplices of the lattice, which is possible whenever $N$ is equal to the number of sites of
the simplices. Based on these results we also expect a two-fold degenerate simplex solid ground state for the \su{4} model on
the pyrochlore lattice (as suggested by previous work~\cite{Henley2001,indergand2006,arovas2008}) as well as the \su{3} model
on the Garnet lattice or variants thereof.~\cite{Henley2001,Bergholtz2010}

In principle, it is sufficient that the number of sites be a multiple of $N$ for this scenario to
be realized. Interestingly, the SU(2) Heisenberg model on the checkerboard lattice fulfills this condition, and
its ground state is also spontaneaously quadrumerized.\cite{fouet,canals} However, the strong bonds
form plaquettes on the {\it empty}
squares in that case, and the mechanism is different: the diagonal bonds are instrumental in frustrating the
inter-plaquette coupling.

Another interesting conclusion can be drawn regarding flavor-wave theory. In models investigated so far,
flavor-wave theory (FWT) has always revealed some aspect of the ground state: it led to the correct
ground state for the SU(3) model
on the square\cite{toth} and the triangular\cite{laeuchli} lattices, %and honeycomb\cite{xiang} lattices,
and it gave interesting insight for the \su{4} model on the
square lattice.\cite{corboz11-su4} In the case of \su{3} on kagome reported in this paper, it fails completely: the ground
state is predicted to be an ordered state with a $\sqrt{3}\times \sqrt{3}$ unit cell and no trimerization (all
bond energies remain equal). In fact, it is simply impossible to describe a trimerized state with FWT since
this would required to build an unfrustrated, two-color configuration on a triangle. It is nevertheless
amusing to notice that the ground state degeneracy is already
lifted at the harmonic level, by contrast to the SU(2) case, and that the $\sqrt{3}\times \sqrt{3}$ state
(energy per site of $-2+2/\sqrt{3}\simeq -0.845$) is favored over the $\vec q= \vec 0$ state (energy per site of 
$-2+4/\pi\simeq -0.727$), as for the SU(2) case beyond linear spin-wave theory.\cite{chubukov,sachdev,Henley19951693} We note that these energies are not variational and thus cannot be compared with the iPEPS energies.

\acknowledgments

The ED simulations have been performed on machines of the platform "Scientific computing" at the University of
Innsbruck - supported by the BMWF, and the iPEPS simulations on the Brutus cluster at ETH Zurich.
We thank the support of the Swiss National Science Foundation, MaNEP, and the Hungarian OTKA Grant No. K73455.

\bibliographystyle{apsrev4-1}
\bibliography{refs}

\end{document}

\title{
Supplementary material
}

\author{Philippe Corboz}
\affiliation{Theoretische Physik, ETH Z\"urich, CH-8093 Z\"urich, Switzerland}
\author{Karlo Penc}
\affiliation{Institute for Solid State Physics and Optics, Wigner Research
Centre for Physics, Hungarian Academy of Sciences, H-1525 Budapest, P.O.B. 49, Hungary}
\affiliation{Department of Physics, Budapest University of
Technology and Economics and Condensed Matter Research Group of the Hungarian Academy of Sciences, 1111 Budapest, Hungary}
\author{Fr\'ed\'eric Mila}
\affiliation{Institut de th\'eorie des ph\'enom\`enes physiques, \'Ecole Polytechnique F\'ed\'erale de Lausanne, CH-1015 Lausanne, Switzerland}
\author{Andreas M. L\"auchli}
\affiliation{Institut f\"ur Theoretische Physik, Universit\"at Innsbruck, A-6020 Innsbruck, Austria}
\affiliation{Max-Planck-Institut f\"{u}r Physik komplexer Systeme, N\"{o}thnitzer Stra{\ss}e 38, D-01187 Dresden, Germany}

\date{\today}

\maketitle
					
\section{Details on the Setup 2 to simulate the kagome model}
Here we explain how to map a nearest-neighbor Hamiltonian on the kagome lattice $\hat h$ onto a nearest-neighbor Hamiltonian on the square lattice $\hat H$, where three physical sites $A$, $B$, $C$ on the kagome lattice are blocked into one block site of the square lattice, as shown in Fig. 2(b) in the main text. The Hilbert space of a block site ${\cal H}_{BS}$ is given by  
\begin{equation}
{\cal H}_{BS} = {\cal H}_{A} \otimes  {\cal H}_{B} \otimes  {\cal H}_{C} ,
\end{equation}
where ${\cal H}_{A}$, ${\cal H}_{B}$, ${\cal H}_{C}$ are the local Hilbert spaces of the three physical sites $A$, $B$, $C$, respectively. Thus, for the SU(3) Heisenberg model with local physical dimension $3$ the local dimension of a block site is $3^3=27$.

The original Hamiltonian defined on the kagome lattice is given by
\begin{equation}
\label{eq:hkag}
\hat h = \sum_{\langle u,v \rangle} \hat h_{u,v},
\end{equation}
where the sum $\langle u,v \rangle$ goes over nearest-neighbor sites on the kagome lattice.  The Hamiltonian $\hat H$ defined on the lattice formed by the block sites reads
\begin{equation}
\hat H = \sum_i  \hat H^s_{i} + \sum_{\langle i,j \rangle_x} \hat H^x_{i,j} + \sum_{\langle i,j \rangle_y} \hat H^y_{i,j} 
 \label{eq:Hsq}
\end{equation}
where the sum $i$ goes over all block sites in the square lattice, and ${\langle i,j \rangle_x}$ and ${\langle i,j \rangle_y}$  are sums over nearest-neighbor pairs in $x$-direction and $y$-direction, respectively. Each site term $\hat H^s_{i}$ contains two  terms of the original Hamiltonian \eqref{eq:hkag}:
\begin{equation}
 \hat H^s_{i}  = \hat h_{i_A, i_B} + \hat h_{i_A, i_C},
\end{equation}
where $h_{i_A, i_B}$ ($\hat h_{i_A, i_C}$) is the Hamiltonian term connecting the physical sites $A$ and $B$ ( $A$ and $C$) on  block site $i$. These two terms correspond to the horizontal and vertical blue lines within a block site in Fig.~2(b).
In $x$-direction the block Hamiltonian is given by the sum of two contributions,
\begin{equation}
\hat H^x_{i,j}  = \hat h_{i_B, j_A} + \hat h_{i_B, j_C},
\end{equation}
where $i_B$ is the $B$-site on block site $i$, and $j_A$ is the $A$-site of the neighboring block site $j$. The first (second) term corresponds to the horizontal (diagonal) blue line in Fig.~2(b) which connect two neighboring block sites in $x$-direction.
Similarly, the block Hamiltonian term in $y$-direction reads
\begin{equation}
 \hat H^y_{i,j}  = \hat h_{i_C, j_A} + \hat h_{i_C, j_B}
\end{equation}
where the first (second) term corresponds to the vertical (diagonal) blue line in Fig.~2(b) which connect two neighboring block sites in $y$-direction. 

Thus, we find that the Hamiltonian \eqref{eq:Hsq} defined on the block site lattice is a nearest-neighbor Hamiltonian on the square lattice, which can be simulated with the standard iPEPS method for the square lattice.  

We point out that the correlations within a block site are taken into account in an exact way, whereas the correlations between physical sites on neighboring block sites depend on the bond dimension $D$. Thus, by performing the blocking we favor correlations between the three physical sites in a block site, which can bias the result for small bond dimensions.  However, as we note in the paper, we do not block the three sites which eventually form a singlet (if we would do this we would bias the state to a singlet state from the start, i.e. a trivial $D=1$ PEPS could already represent a trimerized state). This bias becomes irrelevant for large bond dimensions, because for large $D$ the physical sites on neighboring block sites can also become strongly correlated. Indeed, as we show in the paper, at large bond dimensions the results become equivalent with the ones obtained with the simulation setup I, where we did not use this blocking.